\documentclass[12pt]{article}
\usepackage{graphicx}
\usepackage{amsmath,amssymb}

\newcommand\pubnumber{DPF2013-216}
\newcommand\pubdate{\today}

\def\WSU{Department of Physics and Astronomy\\
Wayne State University, Detroit, MI 48201, USA\\
\vspace{5mm}
		E-mail:  dmitry.zhuridov@wayne.edu}

\def\Title#1{\begin{center} {\Large #1 } \end{center}}
\def\Author#1{\begin{center}{ \sc #1} \end{center}}
\def\Address#1{\begin{center}{ \it #1} \end{center}}

\newcommand\pubblock{\rightline{\begin{tabular}{l} \pubnumber\\
         \pubdate  \end{tabular}}}
\newenvironment{Abstract}{\begin{quotation}  }{\end{quotation}}
\newenvironment{Presented}{\begin{quotation} \begin{center} 
             PRESENTED AT\end{center}\bigskip 
      \begin{center}\begin{large}}{\end{large}\end{center} \end{quotation}}

\def\beq{\begin{equation}}
\def\eeq#1{\label{#1}\end{equation}}
\def\eeqn{\end{equation}}

\def\beqa{\begin{eqnarray}}
\def\eeqa#1{\label{#1}\end{eqnarray}}
\def\eeqan{\end{eqnarray}}

\let\bar=\overbar

\def\Dslash{\not{\hbox{\kern-4pt $D$}}}
\def\dslash{\not{\hbox{\kern-2pt $\del$}}}

\def\msb{{\bar{\ssstyle M \kern -1pt S}}}

\begin{document}
\begin{titlepage}
\pubblock

\vfill
\Title{New Results on Neutrino Magnetic Moments and on Democratic Neutrinos}
\vfill
\Author{ Dmitry Zhuridov\footnote{Work supported in part by the U.S. Department of Energy 
          under contract $\text{DE-SC}0007983$.}} 
\Address{\WSU}
\vfill
\begin{Abstract}
I discuss the two separate issues on neutrino physics. First, the new bounds on tensorial couplings of neutrinos to charged fermions from the existing limits on neutrino transition magnetic moments. Second, explanation of the atmospheric and solar neutrino data in the democratic neutrino theory with only one free parameter (in the leading order), using the effect of incoherence.
\end{Abstract}
\vfill
\begin{Presented}
DPF 2013\\
The Meeting of the American Physical Society\\
Division of Particles and Fields\\
Santa Cruz, California, August 13--17, 2013\\
\end{Presented}
\vfill
\end{titlepage}
\def\thefootnote{\fnsymbol{footnote}}
\setcounter{footnote}{0}

\section{Introduction}

The neutrinos were proposed by W. Pauli \index{Pauli} in 1930~\cite{Pauli:2000ak} and first detected by C.~
Cowan and F.~
Reines in 1956~\cite{Cowan:1992xc}. 
However number of questions on their nature and properties remain unknown~\cite{PDG2012}, e.g., absolute scale and type (Dirac or Majorana) of their masses. 

There are two ways of searching for the true knowledge in particle physics. First, we can generically parametrize possible new physics effects and try to find some correlations with present experimental anomalies. Second, one may reanalyze basics of the underlying theory, which may help to reject number of unphysical results at once. Good example of application of these two methods in astronomy was a historical competition of a developed Ptolemy's and early Copernican models of the solar system. 

In this proceedings I present the two recent researches on the neutrino physics made in both discussed approaches. In the next section I introduce the new constraints on the non-standard neutrino interactions (NSI) from the bounds on neutrino magnetic moments (NMM) derived by K.~Healey, A.~Petrov and DZ~\cite{Healey:2013vka}. In section~\ref{section:democratic_nu} I discuss explanation of the main neutrino experimental results within a new model of democratic neutrinos with significant role of the incoherence of solar neutrinos discovered by DZ~\cite{Zhuridov:2013eqa}.

\section{
Neutrino Magnetic Moment from Nonstandard Interactions}

\subsection{Neutrino Magnetic Moment}

NMM $\mu_{\alpha\beta}$ can be defined by the Hermitian form factor 
${f^M_{\alpha\beta}(0) \equiv \mu_{\alpha\beta}}$ of the term~\cite{Broggini:2012df}
\begin{eqnarray}
	-f^M_{\alpha\beta}(q^2)~   \bar u_\beta(p_2) \, i\sigma_{\mu\nu} q^\nu  u_\alpha(p_1)
\end{eqnarray}
in the 
effective neutrino electromagnetic current
\begin{eqnarray}
	\langle \nu_\beta(p_2) | j_\mu^\text{eff}(0) | \nu_\alpha \rangle 	=	\bar u_\beta(p_2) \, \Lambda_\mu(p_2,p_1)  u_\alpha(p_1),
\end{eqnarray}
where $\alpha,\beta=e,\mu,\tau$ are the flavor indices, $u$ are the spinors, and $q=p_2-p_1$.

In the Standard Model (SM), minimally extended to include Dirac neutrino masses, NMM is suppressed by small masses $m_i$ of 
observable neutrinos~\cite{PDG2012}. 
The diagonal and transition magnetic moments are calculated in the SM to 
be~\cite{Broggini:2012df,Lee:1977tib,Marciano:1977wx,Petcov:1976ff}
\begin{eqnarray}\label{eq:NMMlimitSM}
		\mu_{ii}^\text{D} \approx 3.2\times10^{-20} ~\left( \frac{m_i}{0.1~\text{eV}} \right)~\mu_B
\end{eqnarray}
and
\begin{eqnarray}
		\mu_{ij}^\text{D} 	
		\approx	-4\times10^{-24} ~\left( \frac{m_i + m _j}{0.1~\text{eV}} \right)	\sum_{\ell=e,\mu,\tau} \left(\frac{m_\ell}{m_\tau}\right)^2 U_{\ell i}^*U_{\ell j}	~\mu_B,  \label{eq:NMMlimitSMtransit}
\end{eqnarray}
respectively, where $\mu_B=e/(2m_e)=5.788 \times10^{-5}$ eV\,T$^{-1}$ is the Bohr magneton, and $U_{\ell i}$ is the 
leptonic mixing matrix. In case of Majorana neutrinos
\begin{eqnarray}\label{eq:NMM-Majorana-Dirac}
\mu_{ij}^\text{M}=2\mu_{ij}^\text{D} \quad (\mu_{ij}^\text{M}=0)
\end{eqnarray}
for the opposite (same) $CP$ phases of $i$th and $j$th neutrino mass states.

The best bound on NMM, derived from globular cluster red giants energy loss~\cite{Raffelt:1999gv},
\begin{eqnarray}\label{eq:NMMlimit}
	\mu_\nu < 3\times10^{-12}~\mu_B
\end{eqnarray}
 is far from the SM value. 
The best present laboratory constraint on NMM
\begin{eqnarray}\label{eq:NMMlimitGEMMA}
	\mu_{\bar\nu_e} < 2.9\times10^{-11}~\mu_B  	\qquad	(90\%~\rm{C.L.})
\end{eqnarray}
was obtained in $\bar\nu_e$--$e$ elastic scattering experiment GEMMA~\cite{Beda:2012zz}.

NMM generically induces a radiative correction to the neutrino mass, which constrains NMM~\cite{Bell:2005kz,Bell:2006wi}. 
In the case of diagonal NMM, which is possible only for Dirac neutrinos, the correspondent bound, 
	$\mu_{\alpha\alpha} \lesssim 10^{-14}~\mu_B$, 
significantly strengthens Eq.~\eqref{eq:NMMlimit}. 
However, the transition NMM, which is possible for both Dirac and Majorana neutrino types, is antisymmetric in the flavor indices, and may be suppressed by the SM Yukawas etc., which gives much weaker bound of   $\mu_{\alpha\beta} \lesssim 10^{-9}~\mu_B$~\cite{Bell:2006wi}. 

\subsection{Nonstandard Neutrino Interactions}

There have been many analyses of NSI in neutrino oscillations and neutrino-nucleus scattering 
experiments~\cite{Lee:1977tib,Grossman:1995wx,Bergmann:1999rz,Kopp:2007ne,Barranco:2011wx,Ohlsson:2012kf,Rashed:2013dba}. 
Recently the possibility to constrain NSI, using the existing bounds on transition NMM, was pointed out~\cite{Healey:2013vka}.

Provided that the scale of new physics $M$ is large compared to the electroweak one, NSI of $\nu\nu ff$ type at low-energy scales can be written as~\cite{Bergmann:1999rz,Kopp:2007ne,Ohlsson:2012kf,Rashed:2013dba,Kingsley:1974kq,Cho:1976um,Kayser:1979mj}
\begin{eqnarray}\label{EffLgr}
-\mathcal{L}_\text{eff}= \sum_a
\frac{\epsilon_{\alpha\beta}^{fa}}{M^2} (\bar\nu_\beta \Gamma_a \nu_\alpha) (\bar f \Gamma_a f) 															+	{\rm H.c.},
\end{eqnarray}
where $\epsilon_{\alpha\beta}^{fa}$ are NSI couplings, $f$ denotes the component of an arbitrary weak doublet, $\Gamma_a=\{I,\gamma_5,\gamma_\mu,\gamma_\mu \gamma_5,\sigma_{\mu\nu}\}$, 
$a=\{S,P,V,A,T\}$ and $\sigma_{\mu\nu} = i[\gamma_\mu,\gamma_\nu]/2$. 
Typically only left-handed neutrinos are considered in the literature. This chirality constraint that allows $\nu\nu ff$ interaction only of (axial)vector type does not describe possible leading NSI contribution to NMM. 
However a tensor term in Eq.~(\ref{EffLgr}) can be generated by the Fierz transformation of the scalar terms among the effective low-energy operator set
\begin{eqnarray}\label{EffLgr2}
		\sum_a
\frac{\tilde\epsilon_{\alpha\beta}^{fa}}{M^2} (\bar\nu_\beta \Gamma_a f) (\bar f \Gamma_a  \nu_\alpha) 															+	{\rm H.c.},
\end{eqnarray}
which is presented in 
models with scalar leptoquarks~\cite{Povarov:2007zz}, R-parity-violating supersymmetry~\cite{Gozdz:2012xw}, etc.

\subsection{Theoretical results}

We have found the lowest-order NSI contributions to the transition NMM, using the generic $\nu\nu ff$ parametrization in Eq.~\eqref{EffLgr}.  %
In particular, the operator 
\begin{eqnarray}
	\frac{\epsilon_{\alpha\beta}^{q}}{M^2} (\bar\nu_\beta \sigma_{\mu\nu} \nu_\alpha) (\bar q \sigma^{\mu\nu} q), 
\end{eqnarray}
where $\epsilon_{\alpha\beta}^{q}\equiv\epsilon_{\alpha\beta}^{qT}$ is real, generates NMM
\begin{eqnarray}\label{eq:NMMq}
	\mu_{\alpha\beta} =	\mu_{\alpha\beta}^0 - \sum_q \epsilon_{\alpha\beta}^q \frac{N_cQ_q}{\pi^2}  \frac{m_em_q}{M^2} \,  
	 \ln \left(\frac{M^2}{m_q^2}\right) \mu_B,
\end{eqnarray}
where $N_c=3$ is the number of colors, $Q_q$ is the electric charge of the quark, and $\mu_{\alpha\beta}^0$ denotes the subleading part of NMM that is not enhanced by the large logarithm. 
Eq.~\eqref{eq:NMMq} reproduces the leading order in the exact result, which can be derived in the model with scalar LQs; 
see Ref.~\cite{Povarov:2007zz} for the exact expressions on diagonal NMM. 
Similarly, for the interactions of neutrinos with charged leptons $\ell$, 
\begin{eqnarray}
	\frac{\epsilon_{\alpha\beta}^{\ell}}{M^2} (\bar\nu_\beta \sigma_{\mu\nu} \nu_\alpha) (\bar \ell \sigma^{\mu\nu} \ell),
\end{eqnarray}
\begin{eqnarray}\label{eq:NMMell}
	\mu_{\alpha\beta} =	\mu_{\alpha\beta}^0 +  \sum_\ell \frac{\epsilon_{\alpha\beta}^\ell}{\pi^2}  \frac{m_em_\ell}{M^2} \,   \ln \left(\frac{M^2}{m_\ell^2}\right)  \mu_B
\end{eqnarray}
with $\epsilon_{\alpha\beta}^\ell  \equiv \epsilon_{\alpha\beta}^{\ell T}$. We notice that the dominant logarithmic terms may not contribute to NMM in certain models, e.g., in the SM, due to a mutual compensation between the relevant diagrams~\cite{Petcov:1976ff}. 
For the new physics scale $M=1$~TeV, using Eq.~\eqref{eq:NMMlimit} and taking one nonzero $\epsilon^f_{\alpha\beta}$ at a time, we obtain the constraints shown in Table~\ref{Tab:bounds}.
\begin{table}[htdp]
\caption{Upper bounds on the couplings $\epsilon^f_{\alpha\beta}$.}
\begin{center}
\begin{tabular}{|c|c||c|c||c|c|}
	\hline
	$|\epsilon_{\alpha\beta}^\ell|$	&	Upper bound	&	$|\epsilon_{\alpha\beta}^q|$	&	Upper bound	&	$|\epsilon_{\alpha\beta}^q|$	&	Upper bound 	\\
	\hline
	\hline
	$|\epsilon_{\alpha\beta}^e|$	&	$3.9$	&	$|\epsilon_{\alpha\beta}^d|$	&	$0.49$	&	$|\epsilon_{\alpha\beta}^u|$	&	$0.49$ 	\\
	\hline
	$|\epsilon_{\alpha\beta}^\mu|$	&	$3.0\times 10^{-2}$	&	$|\epsilon_{\alpha\beta}^s|$	&	$3.3\times 10^{-2}$	&	$|\epsilon_{\alpha\beta}^c|$	&	$1.7\times 10^{-3}$ 	\\
	\hline
	$|\epsilon_{\alpha\beta}^\tau|$	&	$2.6\times 10^{-3}$	&	$|\epsilon_{\alpha\beta}^b|$	&	$1.2\times 10^{-3}$	&	$|\epsilon_{\alpha\beta}^t|$	&	$4.8\times 10^{-5}$ 	\\
	\hline
\end{tabular}
\end{center}
\label{Tab:bounds}
\end{table}%

The neutrino-electron and neutrino-nucleus scattering also may constrain the tensorial NSI~\cite{Barranco:2011wx}. 
Using the cross section for the $\bar\nu_e$--$e$ scattering published by the TEXONO Collaboration~\cite{Deniz:2010mp} and taking $M=1$~TeV, 
the bound $|\epsilon^{e}_{e\beta}| < 6.6$ at 90$\%$ C.L. can be obtained~\cite{Barranco:2011wx}, and for the GEMMA sensitivity in Eq.~\eqref{eq:NMMlimitGEMMA} we have
\begin{eqnarray}
	|\epsilon^{e}_{e\beta}| < 2.7  	\qquad	(90\%~\rm{C.L.}),
\end{eqnarray}
which slightly improves the respective bound from NMM. 
The planned $\bar\nu_e$--nucleus coherent scattering experiments can reach the sensitivity of $|\epsilon^{u,d}_{e\beta}|  < 0.2~(M/1\, \rm{TeV})^2$ at 90$\%$ C.L.~\cite{Barranco:2011wx}, which would also improve the respective bounds in Table~\ref{Tab:bounds}.

\section{Democratic Neutrinos and Incoherence}\label{section:democratic_nu}

\subsection{Neutrino Masses and Mixing}

Consider the mass term for three left-handed Majorana neutrinos
\begin{eqnarray}\label{eq:Lm}
	\mathcal{L}_m^\nu	=	-\frac{1}{2}\,	\sum_{\alpha\beta}\bar\nu_{\alpha L}^c M_{\alpha\beta}\nu_{\beta L} + {\rm H.c.},
\end{eqnarray}
where $\alpha,\beta=e,\mu,\tau$ are the flavor indices, $c$ denotes charge conjugation, and
\begin{eqnarray}\label{eq:mass_matrices}
	M	=	 m\left( \begin{array}{ccc}
    0 & 1 & 1 \\ 
    1 & 0 & 1 \\ 
    1 & 1 & 0 \\ 
  \end{array} \right)
\end{eqnarray}
is a ``democratic" mass matrix, which is invariant under the permutation group of three elements $S_3$~\cite{Harari:1978yi,Fritzsch:1995dj,Fukugita:1998vn}. 
The eigenvalues of $M$ result in the mass spectrum\footnote{It is naturally if the degeneracy among the two masses in this spectrum is slightly violated by small perturbations of the matrix $M$.}
\begin{eqnarray}\label{eq:spectrum}  
	\{ m, m, 2m \},
\end{eqnarray}
and the eigenvectors form the mixing matrix of tri-bimaximal~\cite{Fritzsch:1995dj,Wolfenstein:1978uw,Harrison:2002er,Xing:2002sw} {\it type}
\begin{eqnarray}\label{eq:U}
	U	&=&	R_{12}(\theta_{12}) \times R_{23}(\theta_{23})  \nonumber\\
		&=&	 
	\left(   \begin{array}{ccc}
    c_{12} & s_{12}c_{23} & s_{12}s_{23} \\ 
    -s_{12} & c_{12}c_{23} & c_{12}s_{23} \\ 
    0 & -s_{23} & c_{23} \\ 
  \end{array}	\right)	
  		=	\left(	  \begin{array}{ccc}
     \frac{1}{\sqrt{2}} & \frac{1}{\sqrt{6}} & \frac{1}{\sqrt{3}} \\ 
    -\frac{1}{\sqrt{2}} & \frac{1}{\sqrt{6}} &  \frac{1}{\sqrt{3}}  \\ 
    0 & -\frac{2}{\sqrt{6}} & \frac{1}{\sqrt{3}} \\ 
  \end{array}	\right)
\end{eqnarray} 
with $c_{ij}\equiv\cos\theta_{ij}$, $s_{ij}\equiv\sin\theta_{ij}$,  $\theta_{12}=45^\circ$, $\theta_{23}=\pi/2-\arctan(1/\sqrt{2})\approx54.7^\circ$ and nonstandard order of multiplication of the Euler matrixes $R_{ij}$ (compare with Ref.~\cite{Xing:2002sw}).

Note that $U$ is naturally formed by the eigenvectors of $M$ (the eigenvector in the last column of $U$ corresponds to the larger eigenvalue of $M$), and is different from the ordinary tri-bimaximal and ``democratic" mixing patterns (see, e.g., Ref.~\cite{Garg:2013xwa}), which fail to explain the solar neutrino data by the incoherence, as we do in section~\ref{section:solar_nu}.

\subsection[Atmospheric Neutrinos: Oscillations]{Atmospheric Neutrinos: Oscillations}

For $L \ll L^\text{coh}_{ij}$
and
$
	\sigma_x\ll L^\text{osc}_{ij}
$
the neutrino oscillation probability can be written as
\begin{eqnarray}
	P_{\nu_\alpha\to\nu_\beta}{(L,E)}	=	
		\sum_i  |U_{\alpha i}|^2 |U_{\beta i}|^2	+	2\sum_{i>j}  |U_{\alpha i}^*U_{\beta i} U_{\alpha j}U_{\beta j}^*|	\cos\left( \phi_\text{osc} - \phi \right),	
\end{eqnarray}
where $L$ is the base, $L^\text{coh}_{ij}$ ($L^\text{osc}_{ij}$) is the coherence (oscillation) length, $\sigma_x$ is the neutrino wave packet size,  $\phi_\text{osc} = \Delta m_{ij}^2 L / (2E)  =  2\pi L/L^\text{osc}_{ij}$ and $\phi	=	\arg(U_{\alpha i}^*U_{\beta i} U_{\alpha j}U_{\beta j}^*)$.  

For the neutrino masses in Eq.~\eqref{eq:spectrum} and mixing in Eq.~\eqref{eq:U} we have
\begin{eqnarray}\label{eq:PoscModel}
	 P_{\nu_e\to\nu_\tau}{(L,E)}	=	P_{\nu_\mu\to\nu_\tau}{(L,E)}	=	4 s_{12}^2 c_{23}^2	s_{23}^2	\sin^2 \left( 	\frac{\Delta m^2 L}{ 4E}	\right)
							=	\frac{4}{9}		\sin^2 \left( 	\frac{\Delta m^2 L}{ 4E}	\right),  \\
	 P_{\nu_e\to\nu_\mu}{(L,E)}	=								4 c_{12}^2 s_{12}^2	s_{23}^4	\sin^2 \left( 	\frac{\Delta m^2 L}{ 4E}	\right)
							=	\frac{4}{9}		\sin^2 \left( 	\frac{\Delta m^2 L}{ 4E}	\right),   \label{eq:PoscModel2}
\end{eqnarray}
where $\Delta m^2\equiv m_3^2-m_{i<3}^2 =3m^2$. 
Using the atmospheric neutrino mass splitting~\cite{PDG2012}
\begin{eqnarray}
	\Delta m_a^2	=	(\text{2.06--2.67})\times10^{-3}~\text{eV}^2   \qquad   (99.73\%~\text{CL}), 
\end{eqnarray}
the absolute neutrino mass scale can be determined as
\begin{eqnarray}\label{eq:nu_mass_scale}
	0.026~\text{eV}<m<0.030~\text{eV}.
\end{eqnarray}

The difference between the $e$-like and $\mu$-like event distributions in the Super-Kamiokande~\cite{Ashie:2005ik} can be explained using the matter effect on neutrinos which travel through Earth, e.g., 
for the Earth's core electron number density $\bar N^\text{c}_e \approx 5.4~\text{cm}^{-3}~\text{N}_\text{A}$
\begin{eqnarray}
	P^\text{m}_{\nu_e\to\nu_x} (L,E)	\approx	0.05 \,  \sin^2 \left( 	2.8\frac{\Delta m^2 L}{ 4E}	\right), 
\end{eqnarray}
with  $x=\mu,\tau$, which is significantly suppressed with respect to $P_{\nu_\mu\to\nu_\tau}$ in Eq.~\eqref{eq:PoscModel}.

Note that the large amplitude of muon neutrino oscillations is explained from Eqs.~\eqref{eq:PoscModel} and \eqref{eq:PoscModel2} by the muon neutrino oscillations to the electron and tau neutrinos.

\subsection{Solar Neutrinos: Incoherence}\label{section:solar_nu}

Solar $\nu$s are detecting using charged-current (CC) and neutral current (NC) reactions
\begin{eqnarray}
	\nu_e+d	&\to&	e^-+p+p,		\nonumber\\
	\nu_\ell+d	&\to&	\nu_\ell+p+n.	
\end{eqnarray}
Ratio of the neutrino fluxes measured by Sudbury Neutrino Observatory (SNO) with  CC and NC events is~\cite{PDG2012,Aharmim:2005gt}
\begin{eqnarray}\label{eq:CC/NC}
 \frac{\Phi_\text{SNO}^\text{CC}}{\Phi_\text{SNO}^\text{NC}} = \frac{1.68\pm0.06^{+0.08}_{-0.09}}{4.94\pm0.21^{+0.38}_{-0.34}} = 0.340^{+0.074}_{-0.063}.
\end{eqnarray}

For solar neutrinos with the energies $E\lesssim10$~MeV the oscillations due to $\Delta m_a^2$ proceed in the matter of the Sun as in vacuum~\cite{PDG2012}. 
In the natural limit 
\begin{eqnarray}\label{eq:L_coh}
 L^\text{coh}_{ij}=\frac{4\sqrt{2} E^2}{|\Delta m_{ij}^2|} \sigma_x ~ \ll ~ L,	
\end{eqnarray}
where $L\approx 1.5\times10^8~\text{km}$ is the Sun-Earth distance, the oscillation probability takes a simple incoherent form
\begin{eqnarray}\label{eq:incoh}
	P_{\nu_\alpha\to\nu_\beta}^\text{incoh}	=	\sum_i  |U_{\alpha i}|^2 |U_{\beta i}|^2.		
\end{eqnarray}
(Moreover the oscillations due to $\Delta m^2$ should be averaged out already because of the lack of the emitter localization~\cite{Gribov:1968kq,Nussinov:1976uw}.\footnote{The oscillations due to possible violation of the degeneracy in Eq.~\eqref{eq:spectrum}  are suppressed by the solar matter effect.}) 
Using Eq.~\eqref{eq:U}, in perfect agreement with the experimental data in Eq.~\eqref{eq:CC/NC} we have
\begin{eqnarray}
 \frac{\Phi_\text{sol}^\text{CC}}{\Phi_\text{sol}^\text{NC}} =		\frac{P_{\nu_e\to\nu_e}^\text{incoh}}{\sum_\beta P_{\nu_e\to\nu_\beta}^\text{incoh}}	=	 \sum_i|U_{ei}|^4  =	\frac{7}{18}	\simeq	0.39.			
\end{eqnarray}

\subsection{Predictions of Theory with Democratic Neutrinos}

Few basic predictions of the considered theory are as follows 
\begin{itemize}

\item		{\bf Low energy $\boldsymbol\beta$ decays}	\\
			The effective neutrino mass can be calculated using Eqs.~\eqref{eq:U} and \eqref{eq:nu_mass_scale} as
		\begin{eqnarray}
			 \langle m_\beta \rangle 	\equiv	\sqrt{ \sum_i m_i^2 |U_{ei}|^2 }	 =	m\sqrt{2}\approx0.04~\text{eV},
		\end{eqnarray}
			which is much below the KATRIN sensitivity of $0.2$~eV~\cite{Titov:2004pk}, but can be probed by next sub-eV experiments (MARE, ECHO, Project8, etc.).

\item		{\bf Neutrinoless Double Beta Decay} \\ 
			Effective Majorana neutrino mass in the $0\nu2\beta$ decay is vanishing. 

\item		{\bf Neutrino (Transition) Magnetic Moment} \\
Using Eqs.~\eqref{eq:NMMlimitSMtransit}, \eqref{eq:NMM-Majorana-Dirac}, \eqref{eq:spectrum}, \eqref{eq:U} and \eqref{eq:nu_mass_scale}, we have  $\mu_{23}\approx 3.4\times 10^{-24}~\mu_B \gg \mu_{12}, \mu_{13}$.

\end{itemize}

\vspace{6mm}

In conclusion, the two researches in the neutrino physics are presented. First, within a generic parametrization the tensorial nonstandard neutrino interactions are constrained, using the experimental bounds on the neutrino magnetic moments. Second, the basics of the neutrino theory and explanation of the neutrino experiments are reconsidered, and variety of the neutrino data are explained, using the simplest symmetric Lagrangian for the three neutrino species and incoherence of the solar neutrinos at the Earth.



\begin{thebibliography}{99}


\bibitem{Pauli:2000ak} 
  W.~Pauli,
  Camb.\ Monogr.\ Part.\ Phys.\ Nucl.\ Phys.\ Cosmol.\  {\bf 14}, 1 (2000).

\bibitem{Cowan:1992xc} 
  C.~L.~Cowan, F.~Reines, F.~B.~Harrison, H.~W.~Kruse and A.~D.~McGuire,
  Science {\bf 124}, 103 (1956).

\bibitem{PDG2012}
	J. Beringer {\it et al.}  [Particle Data Group], Phys.\ Rev.\ D {\bf 86}, 010001 (2012).

\bibitem{Healey:2013vka} 
  K.~J.~Healey, A.~A.~Petrov and D.~Zhuridov,
  Phys.\ Rev.\ D {\bf 87}, 117301 (2013)
  [arXiv:1305.0584 [hep-ph]].

\bibitem{Zhuridov:2013eqa} 
  D.~Zhuridov,
  arXiv:1304.4870 [hep-ph].


\bibitem{Broggini:2012df} 
  C.~Broggini, C.~Giunti and A.~Studenikin,
  Adv.\ High Energy Phys.\  {\bf 2012}, 459526 (2012)
  [arXiv:1207.3980 [hep-ph]].


\bibitem{Lee:1977tib} 
  B.~W.~Lee and R.~E.~Shrock,
  Phys.\ Rev.\ D {\bf 16}, 1444 (1977).

  
\bibitem{Marciano:1977wx} 
  W.~J.~Marciano and A.~I.~Sanda,
  Phys.\ Lett.\ B {\bf 67}, 303 (1977).

\bibitem{Petcov:1976ff} 
  S.~T.~Petcov,
  Yad.\ Fiz.\  {\bf 25}, 641 (1977)
  [Sov.\ J.\ Nucl.\ Phys.\  {\bf 25}, 340 (1977)]
  Erratum-ibid.\  {\bf 25}, 1336 (1977)
  [Erratum-ibid.\  {\bf 25}, 698 (1977)].

\bibitem{Raffelt:1999gv} 
  G.~G.~Raffelt,
  Phys.\ Rept.\  {\bf 320}, 319 (1999).

\bibitem{Beda:2012zz} 
  A.~G.~Beda {\it et al.}  [GEMMA Collaboration], 
  Adv.\ High Energy Phys.\  {\bf 2012}, 350150 (2012).

\bibitem{Bell:2005kz} 
  N.~F.~Bell, V.~Cirigliano, M.~J.~Ramsey-Musolf, P.~Vogel and M.~B.~Wise,
  Phys.\ Rev.\ Lett.\  {\bf 95}, 151802 (2005)
  [hep-ph/0504134].

\bibitem{Bell:2006wi} 
  N.~F.~Bell, M.~Gorchtein, M.~J.~Ramsey-Musolf, P.~Vogel and P.~Wang,
  Phys.\ Lett.\ B {\bf 642}, 377 (2006)
  [hep-ph/0606248].

 
\bibitem{Grossman:1995wx} 
  Y.~Grossman,
  Phys.\ Lett.\ B {\bf 359}, 141 (1995)
  [hep-ph/9507344].

\bibitem{Bergmann:1999rz} 
  S.~Bergmann, Y.~Grossman and E.~Nardi,
  Phys.\ Rev.\ D {\bf 60}, 093008 (1999)
  [hep-ph/9903517].

\bibitem{Kopp:2007ne} 
  J.~Kopp, M.~Lindner, T.~Ota and J.~Sato,
  Phys.\ Rev.\ D {\bf 77}, 013007 (2008)
  [arXiv:0708.0152 [hep-ph]].

\bibitem{Barranco:2011wx} 
  J.~Barranco, A.~Bolanos, E.~A.~Garces, O.~G.~Miranda and T.~I.~Rashba,
  Int.\ J.\ Mod.\ Phys.\ A {\bf 27}, 1250147 (2012)
  [arXiv:1108.1220 [hep-ph]].


\bibitem{Ohlsson:2012kf} 
  T.~Ohlsson,
  Rep.\  Prog.\  Phys.\  76, {\bf 044201} (2013)
  [Rept.\ Prog.\ Phys.\  {\bf 76}, 044201 (2013)]
  [arXiv:1209.2710 [hep-ph]].
    
\bibitem{Rashed:2013dba} 
  A.~Rashed, P.~Sharma and A.~Datta,
  arXiv:1303.4332 [hep-ph].

\bibitem{Kingsley:1974kq} 
  R.~L.~Kingsley, F.~Wilczek and A.~Zee,
  Phys.\ Rev.\ D {\bf 10}, 2216 (1974).
 
\bibitem{Cho:1976um} 
  C.~F.~Cho and M.~Gourdin,
  Nucl.\ Phys.\ B {\bf 112}, 387 (1976).

\bibitem{Kayser:1979mj} 
  B.~Kayser, E.~Fischbach, S.~P.~Rosen and H.~Spivack,
  Phys.\ Rev.\ D {\bf 20}, 87 (1979).


\bibitem{Povarov:2007zz} 
  A.~V.~Povarov,
  Yad.\ Fiz.\  {\bf 70}, 905 (2007)
  [Phys.\ Atom.\ Nucl.\  {\bf 70}, 871 (2007)].

\bibitem{Gozdz:2012xw} 
  M.~Gozdz,
  Phys.\ Rev.\ D {\bf 85}, 055016 (2012)
  [arXiv:1201.0873 [hep-ph]].


\bibitem{Deniz:2010mp} 
  M.~Deniz {\it et al.}  [TEXONO Collaboration],
  Phys.\ Rev.\ D {\bf 82}, 033004 (2010)
  [arXiv:1006.1947 [hep-ph]].




\bibitem{Harari:1978yi} 
  H.~Harari, H.~Haut and J.~Weyers,
  Phys.\ Lett.\ B {\bf 78}, 459 (1978).

\bibitem{Fritzsch:1995dj} 
  H.~Fritzsch and Z.~-Z.~Xing,
  Phys.\ Lett.\ B {\bf 372}, 265 (1996)
  [hep-ph/9509389].

\bibitem{Fukugita:1998vn} 
  M.~Fukugita, M.~Tanimoto and T.~Yanagida,
  Phys.\ Rev.\ D {\bf 57}, 4429 (1998)
  [hep-ph/9709388].

\bibitem{Wolfenstein:1978uw} 
  L.~Wolfenstein,
  Phys.\ Rev.\ D {\bf 18}, 958 (1978).

\bibitem{Harrison:2002er} 
  P.~F.~Harrison, D.~H.~Perkins and W.~G.~Scott,
  Phys.\ Lett.\ B {\bf 530}, 167 (2002)
  [hep-ph/0202074].

\bibitem{Xing:2002sw} 
  Z.~-Z.~Xing,
  Phys.\ Lett.\ B {\bf 533}, 85 (2002)
  [hep-ph/0204049].

\bibitem{Garg:2013xwa} 
  S.~K.~Garg and S.~Gupta,
  arXiv:1308.3054 [hep-ph].

\bibitem{Ashie:2005ik} 
  Y.~Ashie {\it et al.}  [Super-Kamiokande Collaboration],
  Phys.\ Rev.\ D {\bf 71}, 112005 (2005)
  [hep-ex/0501064].

\bibitem{Aharmim:2005gt} 
  B.~Aharmim {\it et al.}  [SNO Collaboration],
  Phys.\ Rev.\ C {\bf 72}, 055502 (2005)
  [nucl-ex/0502021].

\bibitem{Gribov:1968kq} 
  V.~N.~Gribov and B.~Pontecorvo,
  Phys.\ Lett.\ B {\bf 28}, 493 (1969).

\bibitem{Nussinov:1976uw} 
  S.~Nussinov,
  Phys.\ Lett.\ B {\bf 63}, 201 (1976).

\bibitem{Titov:2004pk} 
  N.~A.~Titov [KATRIN Collaboration],
  Phys.\ Atom.\ Nucl.\  {\bf 67}, 1953 (2004).




\end{thebibliography}
\end{document}